\documentclass[reprint,
 amsmath,amssymb,aip,
]{revtex4-2}
\usepackage{graphicx}
\usepackage{dcolumn}
\usepackage{xcolor}
\usepackage{setspace}
\usepackage{bm}
\usepackage{enumitem}
\usepackage{booktabs}
\usepackage{textcomp}
\usepackage{makecell}
\usepackage{multirow}
\usepackage{wasysym}
\usepackage{xr}
\usepackage[utf8]{inputenc}
\usepackage[T1]{fontenc}
\usepackage{etoolbox}
\usepackage[normalem]{ulem}
\usepackage{comment}
\usepackage{hyperref}
\newcommand{\be}{\begin{equation}}
\newcommand{\ee}{\end{equation}}
\newcommand{\bg}{\begin{equation}}
\newcommand{\eg}{\end{equation}}
\newcommand{\bdm}{\begin{displaymath}}
\newcommand{\edm}{\end{displaymath}}
\newcommand{\bea}{\begin{eqnarray}}
\newcommand{\eea}{\end{eqnarray}}
\newcommand{\beas}{\begin{eqnarray*}}
\newcommand{\eeas}{\end{eqnarray*}}
\newcommand{\ba}{\begin{array}}
\newcommand{\ea}{\end{array}}

\newcommand{\bfg}{\begin{figure}}
\newcommand{\efg}{\end{figure}}

\newtheorem{lm}{Lemma}
\newtheorem{cl}{Corollary}
\newtheorem{df}{Definition}

\newcommand{\blm}{\begin{lm}}
\newcommand{\elm}{\end{lm}}
\newcommand{\bcl}{\begin{cl}}
\newcommand{\ecl}{\end{cl}}
\newcommand{\bdf}{\begin{df}}
\newcommand{\edf}{\end{df}}
\newcommand{\brk}{\begin{rm}}
\newcommand{\erk}{\end{rm}}

\newcommand{\Om}{\Omega}

\newcommand{\ct}{\cite}
\newcommand{\rf}{\ref}

\makeatletter
\def\@email#1#2{%
 \endgroup
 \patchcmd{\titleblock@produce}
  {\frontmatter@RRAPformat}
  {\frontmatter@RRAPformat{\produce@RRAP{*#1\href{mailto:#2}{#2}}}\frontmatter@RRAPformat}
  {}{}
}%
\makeatother
\begin{document}

\title{Experimental demonstration of superdirective spherical dielectric antenna}

\author{Roman Gaponenko}
\email{roman.gaponenko@metalab.ifmo.ru}

\author{Mikhail S. Sidorenko}

\author{Dmitry Zhirihin}
\affiliation{School of Physics and Engineering, ITMO University, 197101, Saint-Petersburg, Russia}

\author{Ilia L. Rasskazov}
\affiliation{KLA Corporation, 5 Technology Drive, Milpitas, California 95035, USA}

\author{Alexander Moroz}
\affiliation{Wave-scattering.com}

\author{Konstantin Ladutenko}

\author{Pavel Belov}

\author{Alexey Shcherbakov}
\affiliation{School of Physics and Engineering, ITMO University, 197101, Saint-Petersburg, Russia}

\date{\today}
\begin{abstract}
An experimental demonstration of directivities exceeding the fundamental Kildal limit, a phenomenon called {\em superdirectivity}, is provided for spherical high-index dielectric antennas with an electric dipole excitation. 
A directivity factor of about 10 with a total efficiency of more than 80\% for an antenna having a size of a third of the wavelength was measured.
High directivities are shown to be associated with constructive interference of particular electric and magnetic modes of an open spherical resonator. 
Both analytic solution for a point dipole and a full-wave rigorous simulation for a realistic dipole antenna were employed for optimization and analysis, yielding an excellent agreement between experimentally measured and numerically predicted directivities.
The use of high-index low-loss ceramics can significantly reduce the physical size of such antennas while maintaining their overall high radiation efficiency. 
Such antennas can be attractive for various high-frequency applications, such as antennas for the Internet of things, smart city systems, 5G network systems, and others. The demonstrated concept can be scaled in frequency.
\end{abstract}
\maketitle

 \section{Introduction}
 \label{sec:intro}
Utilization of dielectrics as high-Q resonators, which gave birth to the first concepts of all-dielectric antennas~\cite{Long1983,McAllister1983,McAllister1984}, has a long history stretching back more than a century ago~\cite{Debye1908, Richtmyer1939, Stratton_1941, Gastine1967}. Such antennas became attractive for various high frequency applications due to numerous advantages, including small physical size, simplicity of design and production, high temperature tolerance, corrosion resistance, high radiation efficiency, wide frequency range, adaptive polarization, ease of integration with other antennas and multiple feed systems, and stable radiation patterns.
This prospect has become appealing with the advent and rapid development of communication networks with significantly increased amount of transmitted data~\cite{5G_tutorial, 6G_tech, lin_2019, 6G_tech2}.
An increased interest in the application of all-dielectric antennas based on spherical resonators is witnessed in numerous theoretical~\cite{Bladel75, Chew1987, moroz_recursive_2005, Evlyukhin2010, Arslanagic2011, Balanis2012, Krasnok2014, Liu2015, Arslanagic2017, Li2018, Rasskazov2019, Wang2019, Majic2020, Gaponenko2021} and experimental~\cite{Wong1993, Filonov2012, Evlyukhin2012, Kuznetsov2012, Geffrin2012, Rolly2013, krasnok2014a, krasnok2014b, Kuznetsov2016, Tribelsky2015, Tribelsky2016, Tsuchimoto2016, Forestiere2020, Sinev2020, Powell2021} works.
Antennas made of materials with simultaneously high refractive index and low losses became increasingly interesting in recent years. 
Due to a large contrast with the environment, such antennas have small radiation leakage, facilitated by high Q-factors of resonances. In the optical band, high-index dielectric nanoparticles have been employed for enhancing the direction-selective absorption and emission of nanoantennas~\cite{Rolly12, ziolk2017, Monti2022}.

The radiation pattern of an antenna is determined by the interference of excited electric and magnetic modes of an open resonator. To obtain a directional design, it is necessary to have constructive mode interference in a given direction and destructive in other directions. The maximum directivity of an antenna is usually considered to be constrained by the so-called Kildal~\cite{kildal_fundamental_2007,kildal_degrees_2017} limit,
\begin{equation}
        \mathcal D_{\rm lim}=\left(kR_{\text{ant}}\right)^{2}+3.
        \label{eq:K-lim}
\end{equation}
The limit is expressed solely in terms of the antenna ``size parameter'', $kR_{\text{ant}}$, where $R_{\text{ant}}$ is the radius of the sphere {\em circumscribing} the antenna and $k$ is the free-space wave number.
Note in passing that this limit does not always work for electrically small dielectric antennas~\cite{bouwkamp_problem_1945,uzkov1946approach, Gaponenko2021}. The antennas that exceed this limit are called \textit{superdirective}.
It is possible to obtain electrically small superdirective antennas~\cite{Newman1982,Yag2012,Haskou2016,Ta2017,Shi22}. Alas such antennas become inefficient~\cite{karlsson_efficiency_2013} as their size decreases, because they require relatively high currents to radiate even low powers. Strong currents lead to an increase in Ohmic losses and large reactive fields near the antenna. Even if Ohmic losses can be significantly reduced by using materials with a low loss angle for the resonator (tan~$\delta<0.001$), the issue of increasing {\em reactive} fields is not so easy to resolve. An increase in the dielectric constant of the resonator with decreasing size leads to a greater localization of the field of \textit{internal} \cite{Wang2019} (or interior \cite{Gastine1967}) modes inside the resonator and to an increase in sensitivity to the slightest changes inside the resonator antennas (e.g., the location and shape of the feeding element).

In view of the above challenges, it has not been obvious that an experimental realization of small superdirective {\em dielectric} resonant antennas with practical parameters were feasible. In this work we demonstrate the existence of such designs via numerical optimization and direct measurements. Moreover, we will show that even very simple designs consisting of a homogeneous dielectric sphere fed by an electric dipole source can meet the superdirectivity requirements while retaining a reasonable bandwidth. In what follows, our theoretical approach, experimental details and measured data are described.

 \section{Methods}
 \label{sec:optim}
 \subsection{Theory}
 \label{sec:optim_sca}
The main parameters characterizing an antenna are the directivity, $\mathcal{D}$, gain, $\mathcal{G}$, and the realized gain, $\mathcal{G}_{\cal R}$.
The {\em directivity} is a dimensionless parameter determined by a relation of the power emitted in some direction to an average of the emitted power over the full solid angle:
\begin{equation}
     \mathcal{D}(\theta, \phi)=\frac{4 \pi |F(\theta, \phi)|^2}{\int_0^{2 \pi}\int_0^{\pi} |F(\theta', \phi')|^2 \sin \theta' \,d\theta'\,d\phi'}\cdot
     \label{eq:directivity}
\end{equation}
Here $|F(\theta, \phi)|$ is the normalized radiation pattern.
$\mathcal{D}$ is determined solely by the shape of the antenna pattern and
does not take into account the antenna radiation efficiency, $e$, nor the reflection losses due to an impedance mismatch.

The {\em gain} of an antenna, $\mathcal{G}$, is related to the directivity by $\mathcal{G}=e_{cd} \mathcal{D}$, where $e_{cd}$ is the antenna radiation efficiency ($0\leq e_{cd}\leq 1$). The {\em realized} gain, $\mathcal{G}_{\cal R}$, is the gain of an antenna reduced by its impedance mismatch factor~\cite{IEEEdefs} and related to the directivity through the total antenna efficiency ($0\leq e_0\leq 1$) by the formula $\mathcal{G}_{\cal R}=e_0 \mathcal{D}$.
In the measurement setup considered in this work (see Fig. \rf{fig:setup}), the gain, $\mathcal{G}$, and realized gain, $\mathcal{G}_{\cal R}$, can be derived from the Friis transmission equation~\cite{Friis1946} for the line-of-sight communication between two antennas in a lossless medium (air)\cite{Balanis2016}:
\begin{equation}
   \mathcal{G}_{\cal R}(\theta,\phi)=\frac{P_{rec}}{P_{tr}}\bigg(\frac{4 \pi r}{\lambda}\bigg)^2 \frac{1}{p\, \mathcal{G}_{rec}(1-|\Gamma_{rec}|^2)},
    \label{eq:realized_gain}
\end{equation}
\begin{equation}
   \mathcal{G}(\theta,\phi)=\mathcal{G}_{\cal R}(\theta,\phi)/e_r=\mathcal{G}_{\cal R}(\theta,\phi)/(1-|\Gamma_{tr}|^2),
    \label{eq:gain}
\end{equation}
where $p$ is the polarization mismatch factor (in our case $p=1$), $e_r$ is the reflection (mismatch) efficiency, $\lambda$ is the free space wavelength, $P_{tr}$ refers to the power provided to a transmission line attached to an emitting antenna, $P_{rec}$ refers to the power received from a transmission line attached to a receiving antenna. $\Gamma$ is the reflection coefficient for impedance mismatch between the antenna and a transmission line. Subscripts `$tr$' and `$rec$' are used here for transmitting antenna (superdirective antenna under study) and the receiving horn antenna. $\mathcal{G}_{rec}$ is the gain of the receiving horn antenna in the direction strictly oriented to the transmitting antenna (measured by the two-antenna method~\cite{Ant_meas}).

A general physical picture of light scattering by a spherical particle under plane wave illumination is well described by the Lorenz–Mie theory \cite{Bohren1998}, where the solution is represented as an infinite multipole series of partial vector spherical waves. The theory provides an exact solution regardless of the wavelength and the size of the sphere.
The Lorenz–Mie  theory was later generalized to the case of a point dipole excitation source~\cite{moroz_recursive_2005}. Note in passing that such an analytical solution is only available for perfectly spherical surfaces. Any other geometric shapes require more sophisticated numerical methods~\ct{Waterman1971,Mishchenko_2014}.

The physics underlying resonant scattering behavior of a dielectric resonator is the excitation of its electrical and magnetic modes. Resonant modes of an open spherical resonator can be found from exact analytical formulas for a homogeneous sphere~\cite{Gastine1967, Bohren1998}. For the respective TE$_{\ell mq}$ and TM$_{\ell mq}$ modes these formulas are
\begin{align}
\begin{aligned}
    \frac{j_{\ell-1}(\eta_1 kR_1)}{j_{\ell}(\eta_1 kR_1)}&=\frac{h^{(2)}_{\ell-1}(kR_1)}{\eta_1 h^{(2)}_{\ell}(kR_1)}, 
    \\
    \frac{\ell}{\eta_1 kR_1}-\frac{j_{\ell-1}(\eta_1 kR_1)}{j_{\ell}(\eta_1 kR_1)}&=\frac{\eta_1 \ell}{kR_1}-\eta_1\frac{h^{(2)}_{\ell-1}(kR_1)}{h^{(2)}_{\ell}(kR_1)}\cdot \\ 
    \label{eq:char_eq}
\end{aligned}
\end{align}
Here $j_{\ell}$ and $h^{(2)}_{\ell}$ are the usual spherical Bessel function of the first kind and the spherical Hankel function of the second kind~\cite{Abramowitz1964}, respectively, $\ell$ stands for multipole order, index $q$ labels the subsequent roots of Eqs.~(\ref{eq:char_eq}), $\eta_1=\sqrt{\varepsilon_1}$ is the refractive index of the sphere material, $R_1$ is the radius of the homogeneous sphere, and $k$ is a free space wavenumber.
The equations above are independent of the azimuthal mode number $m$ due to the spherical symmetry. All $2\ell+1$ modes with different $m$ are degenerate in frequency~\cite{Gastine1967}.

\begin{table}[b!]
\caption{\label{table:rescon}Zeros $\zeta_{\ell q}$ of the spherical Bessel function of the first kind $j_{\ell}(\zeta_{\ell q}=\eta_1 k R_1)=0$. The subscript $q$ denotes here the ordinal number of the zero of the $\ell$-th order spherical Bessel function.}
\begin{ruledtabular}
\begin{tabular}{c|cccc}
       &    $q=1$  &   $q=2$   &    $q=3$  &   $q=4$\\
\hline
 $\ell=0$ & $3.14159$ & $6.28319$ & $9.42478$ & $12.5664$\\
 $\ell=1$ & $4.49341$ & $7.72525$ & $10.9041$ & $14.0662$\\
 $\ell=2$ & $5.76346$ & $9.09501$ & $12.3229$ & $15.5146$\\
 $\ell=3$ & $6.98793$ & $10.4171$ & $13.698$  & $16.9236$\\
 $\ell=4$ & $8.18256$ & $11.7049$ & $15.0397$ & $18.3013$\\
 $\ell=5$ & $9.35581$ & $12.9665$ & $16.3547$ & $19.6532$\\
\end{tabular}
\end{ruledtabular}
\end{table}

For spherical resonators with high refractive index, approximate conditions for excitation of the electric TM$_{\ell mq}$ and magnetic TE$_{(\ell+1)mq}$ modes are closely related. They can be roughly estimated in the limit $\eta_1\rightarrow\infty$, in which case Eqs.~(\ref{eq:char_eq}) reduce to
\begin{equation}
    j_{\ell}(\zeta_{\ell q}) \simeq 0,
    \label{eq:char_eq_short}
\end{equation} 
with $\zeta_{\ell q}=\eta_1 k R_1$ being the $q$-th zero of the $\ell$-th order spherical Bessel functions of the first kind $j_{\ell}$ (see Table~\ref{table:rescon}). Approximate Eq.~(\ref{eq:char_eq_short}) can be used for a quick and easy estimate of the frequency position of the relevant resonator modes.

At the first stage of this work, initial designs with high directivity were obtained by performing optimization using the analytical solution for a sphere excited by a point electric dipole~\cite{moroz_recursive_2005}. In order to precisely match the experiment, a subsequent optimization was performed in the CST Studio Suite~\cite{CST} taking into account a {\em finite} and realistic dipole size and connector cables. During optimization, our search for optimal parameters of the dipole was performed by varying (i) the position of the dipole, (ii) the thickness and length of the dipole arms, (iii) their offset and bend radius, (iv) the distance between the dipole arms. A frequency shift of resonance was achieved, if required, by changing either the resonator permittivity or its size. The above approach allowed us to obtain the designs of directional antennas with high total antenna efficiency under realistic conditions, closely matching the experiment.

\begin{figure}[h!]
\begin{centering}
\includegraphics[width=8.4cm]{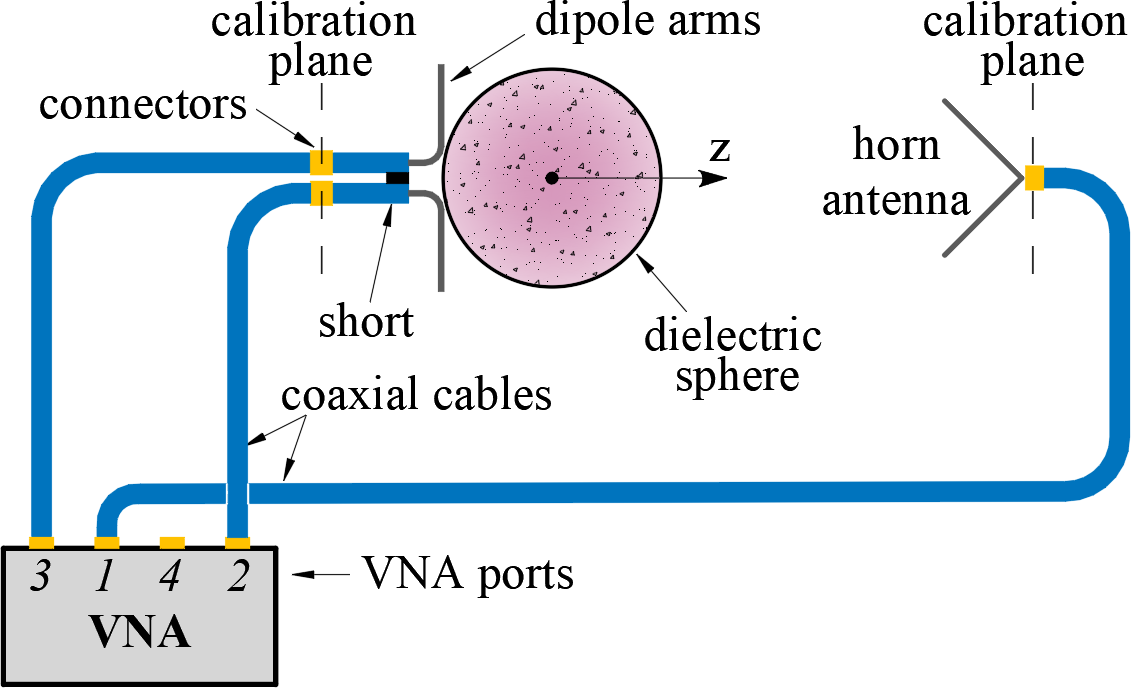}
\par\end{centering}
\caption{\label{fig:setup}The scheme of a three-port connection of an experimental setup for the radiation pattern measurement using R\&S ZVB$20$ Vector Network Analyzer.}
\end{figure}

 \subsection{Experimental setup and measurements}
 \label{sec:results_setup}

Experimental measurements of the directivity were carried out in an anechoic chamber using R\&S$^\text{®}$ZVB$20$ Vector Network Analyzer (VNA). 
The horn antenna was mounted on a triaxial positioner. The antenna under study was located on a special foam pedestal on a rotary positioner (see photographs in Section I of the supplementary materials).
As indicated in Fig.~\ref{fig:setup}, experimental data were obtained using a three-port connection to the VNA by measuring the S-parameters in the planes $\phi=0^{\circ}$ and $\phi=90^{\circ}$ with a $\theta$ angle step of $1^{\circ}$ (see Section~II of the supplementary materials). The secant plane of $\phi=0^\circ$ includes the $z$-axis and the arms of the dipole, while the plane of $\phi=90^\circ$ is perpendicular to the arms of the dipole.

The measurements were carried out within $1.4-2.6$~GHz band with a frequency resolution of $1$~Hz. 
The antenna under study is located in the far field of the horn antenna at a distance of $3$ meters.
To implement an electric dipole, two coaxial cables $RG-58$ $A/U$ with a characteristic impedance of $50\pm 2\, \Om$ located side by side were used. 
The arms of the dipole were made from the inner conductors of coaxial cables with a diameter of $1$ mm.
As illustrated in Fig.~\ref{fig:setup}, the horn antenna was connected to the physical port $1$ of the VNA, which was represented by a logical port $s1$. The physical ports $2$ and $3$ were connected to the dipole arms and combined into a differential logical port $d2$.

The quantity $S_{s1d2}$ describing the transmission from the balanced port with differential mode stimulus $d2$ to the single-ended port $s1$ was measured in the experiment. The radiation pattern $|F(\theta, \phi)|$ in Eq.~(\ref{eq:directivity}) is determined as $|S_{s1d2}(\theta, \phi)|$ divided by $\max(|S_{s1d2}(\theta, \phi)|)$, where the maximum is taken over all possible angular directions. The ratio $P_{rec}/P_{tr}$ in Eq.~(\ref{eq:realized_gain}) is equal to the measured value $|S_{s1d2}|^2$. 
The reflection coefficients for each antenna are $\Gamma_{rec}=S_{s1s1}$ and $\Gamma_{tr}=S_{d2d2}$. In general, all the $S$-parameters can be obtained by measuring reflection and transmission between individual ports in accordance with the theory describing multiport experimental measurements~\cite{Fan2003, Rohde2004}:

\begin{widetext}
\begin{equation}
    S=  \begin{bmatrix}
        S_{s1s1} & S_{s1d2} & S_{s1c2}\\
        S_{d2s1} & S_{d2d2} & S_{d2c2}\\
        S_{c2s1} & S_{c2d2} & S_{c2c2}
        \end{bmatrix}
        =\frac{1}{2}\begin{bmatrix}
        2S_{11} &&& \sqrt{2} (S_{12}-S_{13}) &&& \sqrt{2} (S_{12}+S_{13})\\
        \sqrt{2}(S_{21}-S_{31}) &&& S_{22}-S_{23}-S_{32}+S_{33} &&& S_{22}+S_{23}-S_{32}-S_{33}\\
        \sqrt{2}(S_{21}+S_{31}) &&& S_{22}-S_{23}+S_{32}-S_{33} &&& S_{22}+S_{23}+S_{32}+S_{33}
        \end{bmatrix},
\end{equation}
\end{widetext}
where the subscript 's' stands for the single mode, 'd' - for the differential mode, and 'c' - for the common mode.

\begin{figure}[bh!]
\begin{centering}
\includegraphics[width=8.25cm]{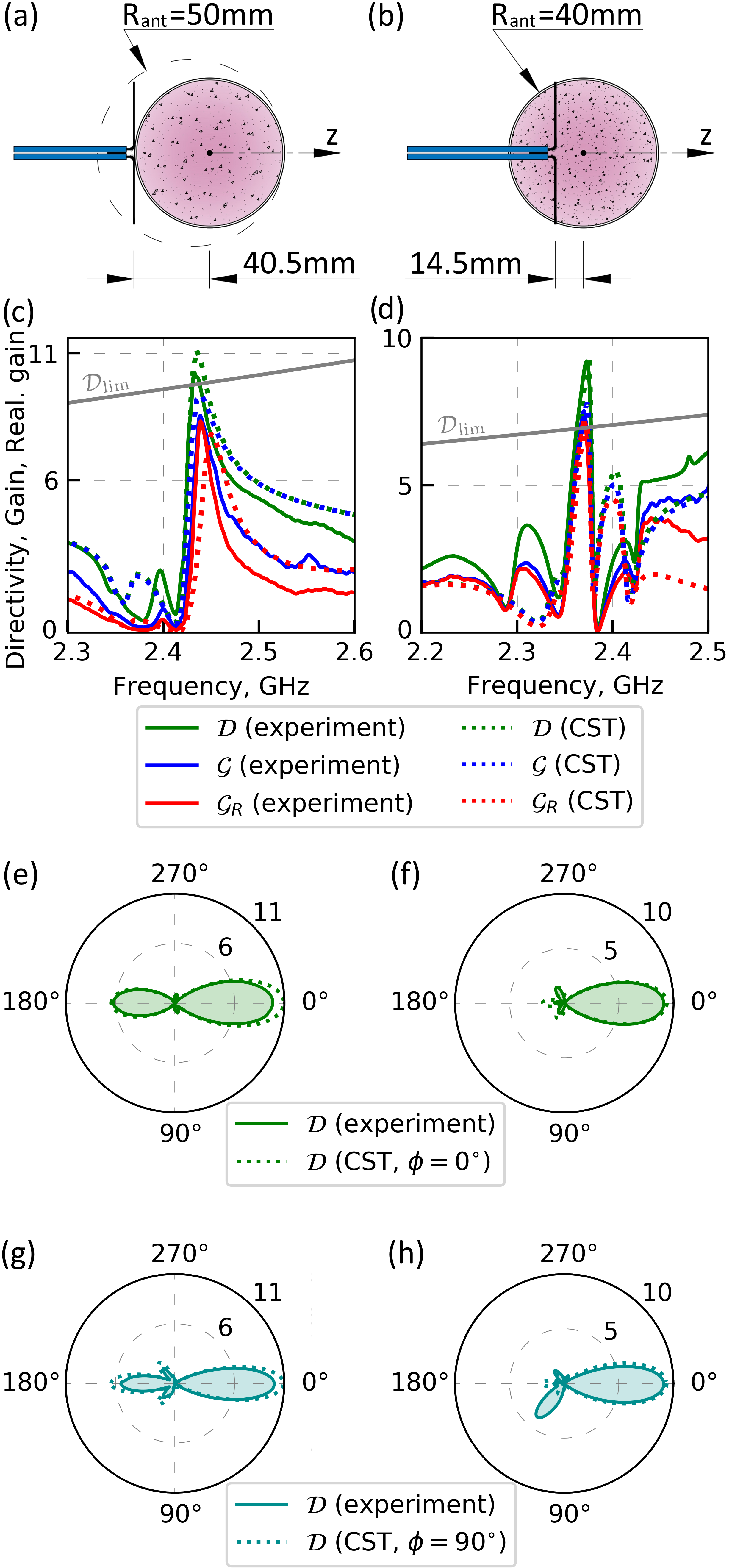}
\par\end{centering}
\caption{\label{fig:powder_sph}(a)-(b) Schematic representation of our antennas in the form of a dielectric sphere excited by an (a) externally and (b) internally placed electric dipole. The radii of circumscribing spheres are $50$ mm and $40$ mm, respectively. (c)-(d) Dependence of directivity $\mathcal{D}$, gain $\mathcal{G}$, and realized gain $\mathcal{G}_{\cal R}$ in the forward direction as a function of frequency for the respective antennas shown in (a) and (b).
(e)-(h) Dependence of the directivity on the angle $\theta$ in the planes $\phi=0^{\circ}$ and $\phi=90^{\circ}$ at the frequencies $2.433$ GHz and $2.373$ GHz (in linear scale) corresponding to the maximum values of directivity of the antenna in the respective column. Numbers on concentric circles in the polar plots reflect corresponding directivity values.}
\end{figure}
\begin{figure}[bt!]
\begin{centering}
\includegraphics[width=8.4cm]{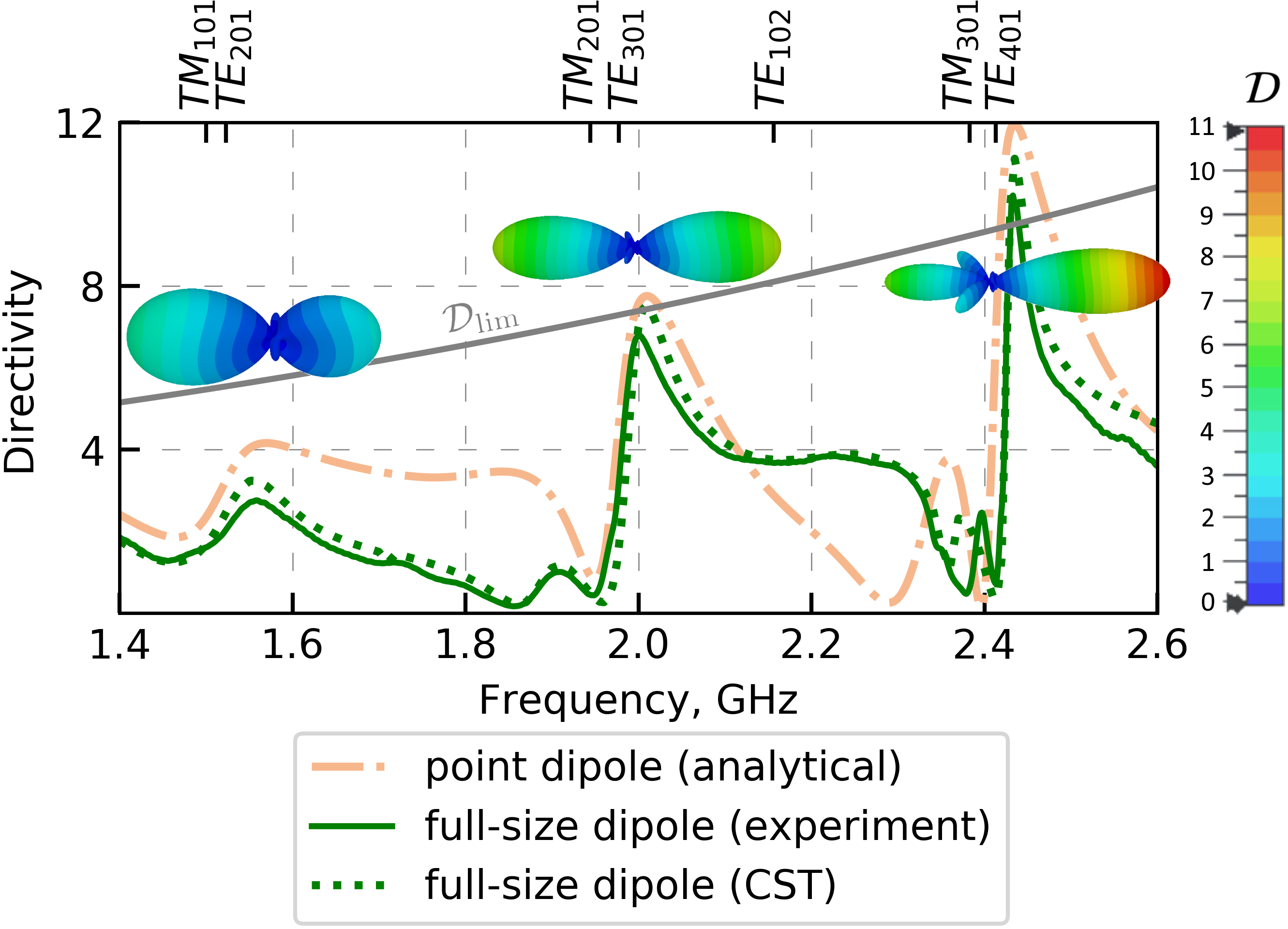}
\par\end{centering}
\caption{\label{fig:freq}Spectra of the directivity ($\mathcal{D}$) in the forward direction (along the $z$-axis) in the range $1.4-2.6$ GHz for the cases of a full-size and point electric dipole oriented tangentially and located at a distance of $r_d=40.5$ from the center of the spherical resonator with the radius of $40$ mm. Antenna construction is described in Section~\ref{sec:results_dip_out}. Calculations of the model with the point dipole were done with a proprietary code based on the theory developed in Ref.~\citenum{moroz_recursive_2005}.
The receiving horn antenna and the transmitting dipole antenna were located at $3.4$ meters distance.
Three-dimensional antenna patterns were obtained in the CST Studio Suite and correspond to the points of local maxima on the green dotted curve. The angular distribution of the directivity values corresponds to the color bar shown on the right. The ticks at the top of the graph correspond to the eigenmode frequencies of the spherical resonator used. 
The Kildal limit is shown with solid gray line; the minimum radius of the sphere describing the antenna was assumed to be $R_{ant}=50$ mm.}
\end{figure}

 \section{Results}
 \label{sec:results}

 \subsection{External dipole}
 \label{sec:results_dip_out}
In order to achieve the superdirective regime, we first considered a dipole located near a spherical resonator -- a system which might be one of the most interesting from a practical viewpoint and the simplest for an experimental implementation. The dielectric antenna was made in the form of a plastic ball filled with ceramic powder and excited externally by an electric dipole as shown in Fig.~\ref{fig:powder_sph}(a). The length of the dipole arms is $38$ mm, the gap between the centers of the cables supplying the dipole is $\sim 4$ mm. The plastic shell made of acrylic with a thickness of $\sim1.1$ mm has an outer radius $R_2=40$ mm, the real part of permittivity $\varepsilon'_2\simeq1.7$, and the loss tangent $\tan \delta_2\simeq0.02$. 
Free Flowing Dielectric Powder ECCOSTOCK HiK has the real part of permittivity $\varepsilon'_1\simeq 12.2$ and the loss tangent $\tan\delta_1\simeq 0.0007$. Vibrations were used to obtain a dense packaging of the ceramic powder inside the spherical resonator. 

The plastic shell used in this work exclusively to set up the shape of the resonator was taken into account in all our calculations. The thickness of the dielectric shell is $\sim 40$ times less than the thickness of the main layer with dielectric powder. Given that the characteristics of the shell are much closer to air than to the powder, the results with and without the shell appear pretty much the same (see Section~IV of the supplementary materials).

At the first stage, an interval containing the frequencies of the first eigenmodes of a homogeneous spherical resonator was considered. Figure~\ref{fig:freq} shows a comparison of the directivity calculated analytically for a sphere excited by a tangentially oriented point dipole with the results obtained by the full-size simulation in CST Studio Suite and measured experimentally. Experimental results seem to agree well with the simulations. At the same time, the analytical point dipole model~\cite{moroz_recursive_2005} describes quite well the position and magnitude of jumps in the directivity. Such Fano-type~\cite{Tribelsky2016} directivity behavior near resonant frequencies arises due to the interaction of a relatively narrow magnetic mode (of the order of $\ell+1$) with a broad electric mode (of the order of $\ell$).

\begin{figure*}[t!]
\includegraphics[width=\textwidth]{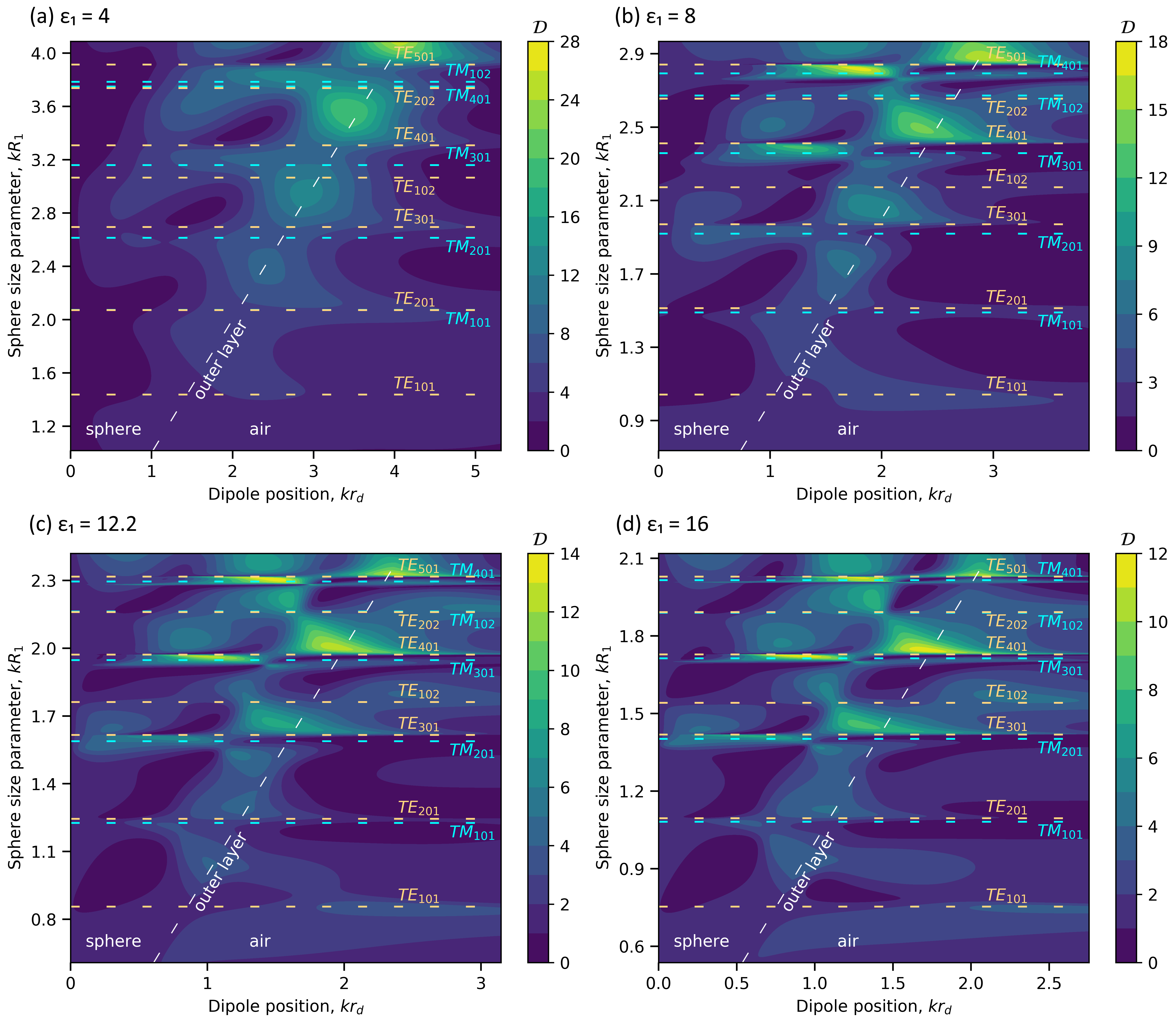}%
\caption{\label{fig:dirmap}
Simulated dependence of the directivity in the forward direction (in linear scale) of a spherical homogeneous dielectric resonator antenna on its dimensionless geometric size and the position of a tangentially oriented point dipole. Maps are presented for spheres with permittivity (a) $\varepsilon_1=4$, (b) $\varepsilon_1=8$, (c) $\varepsilon_1=12.2$, (d) $\varepsilon_1=16$. The position of the electric (TM) and magnetic (TE) eigenmodes of the spherical resonator is shown by dashed bright blue and peach horizontal lines, respectively. The dimensionless values of the sphere size parameter for the presented eigenmodes were obtained by solving Eqs.~(\ref{eq:char_eq}).}
\end{figure*}

Figure~\ref{fig:powder_sph}(c) shows numerical and experimental results for directivity, gain and realized gain in forward direction versus frequency. The data are given for the frequency range $2.3-2.6$ GHz in the vicinity of the directivity resonance.
The experimentally obtained directivity of the antenna reaches $10$ at frequency $2.433$ GHz. The half-power beamwidths (HPBWs) obtained numerically were equal to $47.3^{\circ}$ and $34.1^{\circ}$ for the $\phi=0^{\circ}$ and $\phi=90^{\circ}$ planes, respectively. In the experimental results, HPBWs were $48.6^{\circ}$ and $35.5^{\circ}$ for the indicated planes. There is a reasonable agreement between the experimental data calculated by formulas~(\ref{eq:directivity})--(\ref{eq:gain}) and the simulation results obtained with the CST Studio Suite within the entire demonstrated frequency range. High total antenna efficiency is achieved at an antenna impedance close to $50\Omega$, resulting in minimal energy reflection back to the port and maximum radiation to the environment. The physics underlying the resonant directional behavior for the system under consideration is simultaneous excitation of the electric TM$_{301}$ and magnetic TE$_{401}$ modes of the sphere by the electric dipole and constructive interference of the radiation in the forward direction. This is easy to check, since for these two modes of the spherical resonator the approximate relation $\zeta_{3 1}=\eta_1 k R_1 \simeq 6.988$ is satisfied \big(see Eq.~(\ref{eq:char_eq_short})\big).

Figures~\ref{fig:powder_sph}(e) and \ref{fig:powder_sph}(g) provide a comparison of the experimental and numerical radiation patterns in two planes ($\phi=0^\circ$ and $\phi=90^\circ$) for the resonant frequency of $2.433$ GHz shown in Figure~\ref{fig:powder_sph}(c). 
The difference between the experimentally measured and numerically calculated diagrams in Figs. \ref{fig:powder_sph}(e,g) is associated with the effect of radiation from the coaxial cable feeding the antenna under test due to the currents arising on the cable surface: (i) for $\phi=0^{\circ}$ on the horizontal section of the cable, (ii) for $\phi=90^{\circ}$ on the vertical one. The radiation pattern outside the resonant frequency is affected also by side lobes that arise uncontrollably due to the imperfection of the antenna.

\subsection{Internal dipole}
\label{sec:results_dip_in}
Although the use of a spherical resonator makes it possible to obtain the superdirectional radiation, the matched dipole located outside has a significant length and increases the overall size of the antenna (the circumscribing sphere).
This makes it difficult to exceed Kildal's limit \big(Eq.~(\ref{eq:K-lim})\big) while maintaining high total antenna efficiency. Therefore, we also considered the case of the electric dipole located inside a spherical dielectric resonator. 
The antenna with a dipole inside was made in the form of a plastic ball, similar to the antenna described above, but with an exciting electric dipole located inside, as shown in Fig.~\ref{fig:powder_sph}(b). 
The dipole was fixed in pre-perforated holes at three points of the plastic sheath: at the place where the cable crosses it, and at the ends of the dipole arms. Then, the small gaps remaining in the holes were sealed with a hot glue and after that the powder was poured into the shell.
Fabrication and measurement processes are presented in Section I of the supplementary materials. The receiving horn antenna and the transmitting antenna were located at $2.7$ meters distance in this case.

\begin{figure*}[t!]
\includegraphics[width=\textwidth]{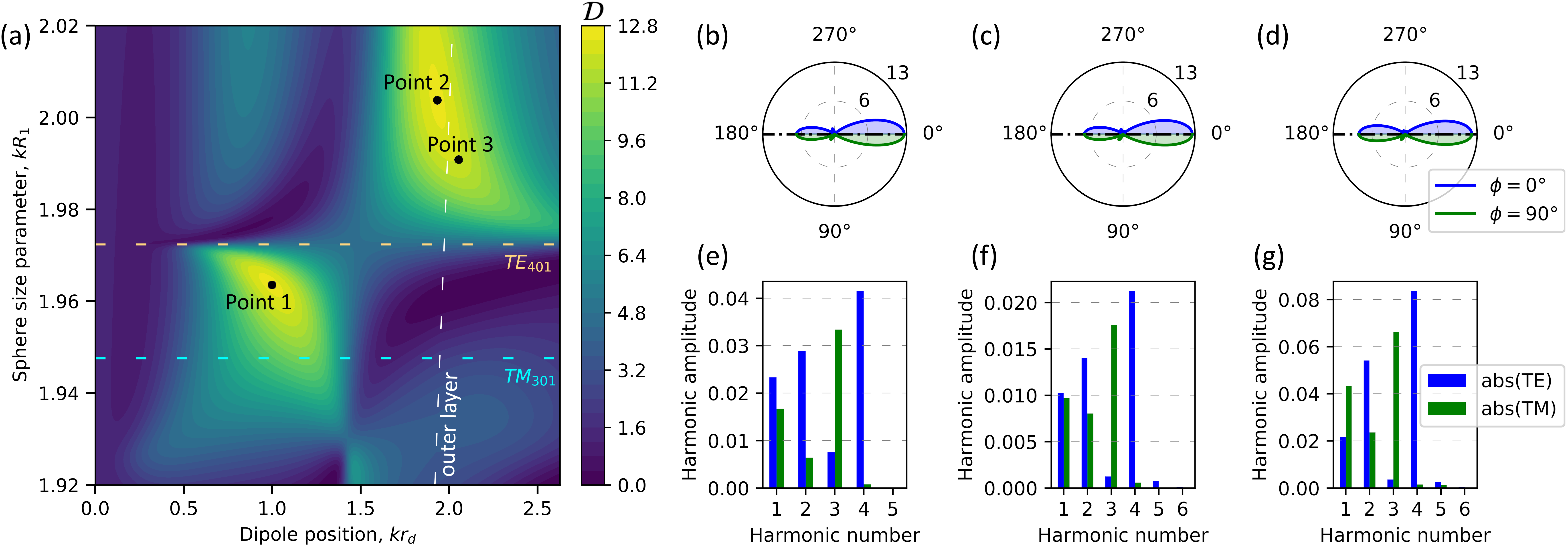}%
\caption{\label{fig:dirmap2}(a) Simulated dependence of the directivity in the forward direction (in linear scale) of a spherical homogeneous dielectric resonator antenna on its dimensionless geometric size and the position of a tangentially oriented point dipole for a fragment of Figure~\ref{fig:dirmap}(c) near the TE$_{401}$ mode resonance. Map is presented for sphere with permittivity $\varepsilon_1=12.2$. The positions of the electric TM$_{301}$ and magnetic TE$_{401}$ eigenmodes of the spherical resonator are shown by dashed bright blue and peach horizontal lines, respectively.
Polar plots (b), (c) and (d) show the directivity versus polar angle $\theta$ for points $1$, $2$, and $3$, respectively, marked on panel (a). The diagrams are axially symmetrical, so two different planes are combined into one graph: the upper half shows the plane containing the dipole vector and the $z$ axis ($\phi=0^\circ$), the lower half describes the perpendicular plane ($\phi=90 ^\circ$). 
Below each polar plot, the corresponding multipole expansion is shown in (e), (f), and (g).
}
\end{figure*}

The \textit{internal} modes of a high-index dielectric spherical resonator are generally more efficiently excited from within its interior~\cite{Gaponenko2021}. 
For initial search of the optimal position of the dipole, a homogeneous spherical antenna excited by a tangentially oriented point dipole was considered. Figure~\ref{fig:dirmap} shows the directivity in the forward direction as a function of the dimensionless parameter of antenna size and dipole position for several values of permittivity.
It is noticeable that with increasing the dielectric constant, the geometric size of the resonator decreases and the quality factor of the resonances increases. Figure~\ref{fig:dirmap2} shows an enlarged fragment of Figure~\ref{fig:dirmap}(c) near the resonant condition of the TE$_{401}$ mode. For the points with the maximum directivity, the dependence of the directivity on the angle $\theta$ in the planes $\phi=0^{\circ}$ and $\phi=90^{\circ}$ are plotted and the corresponding multipole expansions are displayed. It turns out that the ratio of harmonic amplitudes is approximately the same for both the point dipole located inside and at the resonator boundary.
By means of full-size optimization, the optimal position of the dipole is chosen on the $z$-axis and shifted by $14.5$ mm from the center of the sphere, which places the dipole arms near the zone with the highest directivity in Figure~\ref{fig:dirmap2} \big(marked point (a)\big). The length of the dipole arms was $36$ mm, the gap between the centers of the cables was $4$ mm. Such an experimental setup enables one to reduce the radius of the antenna circumscribing sphere down to $40$ mm, i.e. to the size of the dielectric spherical resonator. 

Figure~\ref{fig:powder_sph}(d) shows numerical and experimental results for directivity, gain and realized gain as a function of frequency in the $z$-direction. The data are given for the frequency range $2.2 - 2.5$ GHz in the vicinity of the directivity resonance. The experimentally measured radiation patterns at the resonant frequency of $2.373$ GHz in two planes are shown in Figures~\ref{fig:powder_sph}(f) and \ref{fig:powder_sph}(h). The directivity of the antenna reaches $9.1$ in the direction $\theta=0^{\circ}$. The antenna exceeds Kildal limit while maintaining high total antenna efficiency at the resonant frequency. The HPBWs that were numerically obtained are $45.9^{\circ}$ and $44.5^{\circ}$ for the $\phi=0^{\circ}$ and $\phi=90^{\circ}$ planes, respectively. In the experimental results, HPBWs are $47.5^{\circ}$ and $40.1^{\circ}$ for the indicated planes.

An increase in directivity can also be achieved for an external dipole by immersing the dipole under the outer surface of the resonator \big(point (b) in Figure~\ref{fig:dirmap2}\big).
Using a small notch on the surface to place a dipole inside can also produce super-directional radiation, as has been numerically demonstrated in Ref.~\citenum{krasnok2014a}.

\begin{figure}[b!]
\begin{centering}
\includegraphics[width=0.48\textwidth]{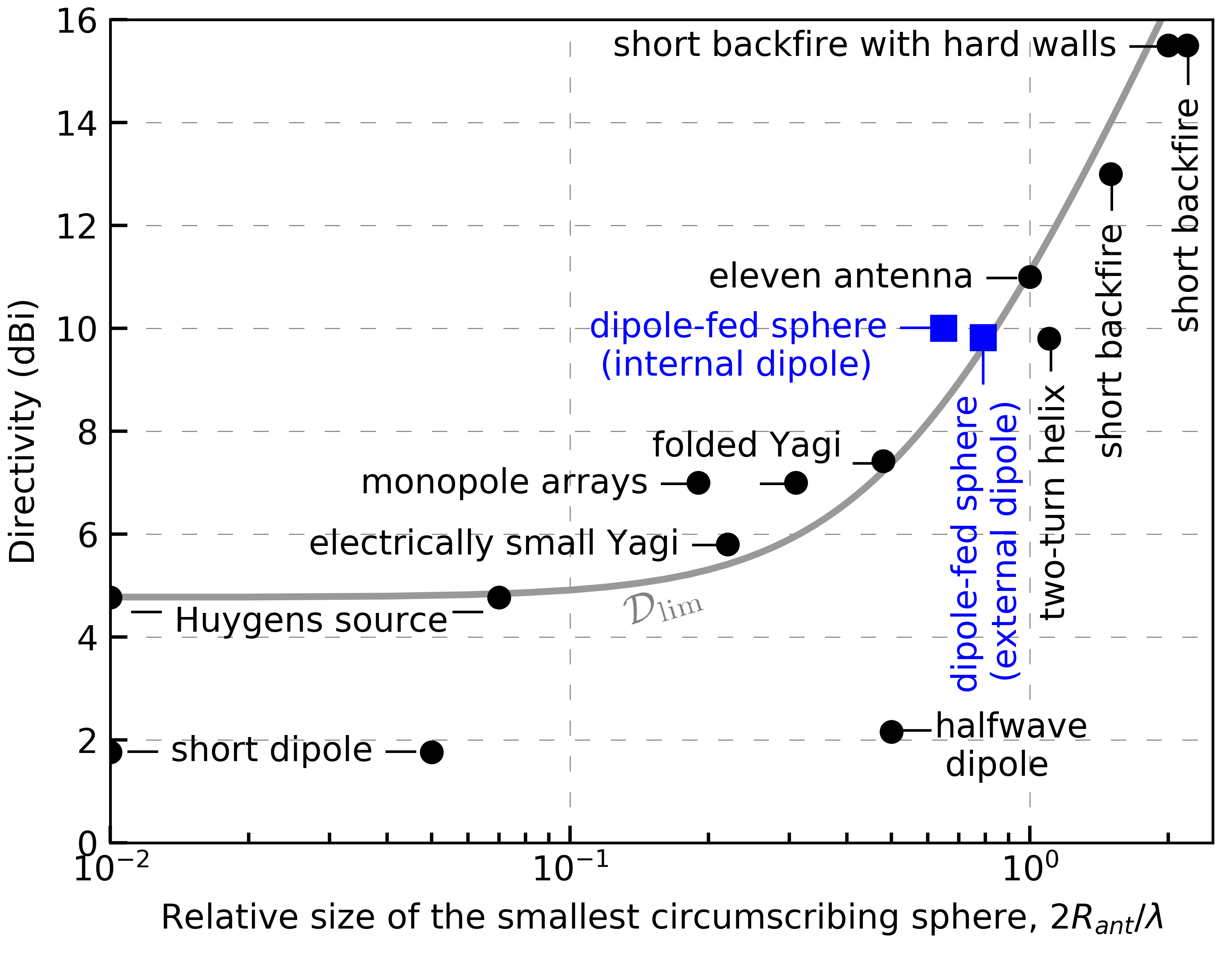}
\par\end{centering}
\caption{Directivity of important antennas demonstrated in the literature (black circles)~\cite{kildal_degrees_2017} and experimentally measured in this work (blue squares).
The Kildal limit is shown with solid gray line.
}
\label{fig:example}
\end{figure}

\section{Discussion}
\label{sec:discussion}
Figure~\ref{fig:example} shows a comparison of our results with available experimental data (Fig.~3 of Ref.~\citenum{kildal_degrees_2017}). Our results unambiguously demonstrate that it is possible to obtain both highly efficient and superdirective radiation simultaneously. Without taking into account the radiation efficiency it is possible to arrive at even higher directivity values by about $15\%$ for the same homogeneous sphere at a given resonance. 

In view of potential applications, an assessment of antenna tolerances and possible sources of errors for both numerical and experimental data deserves particular attention. The analytical solution of an ideal point-like dipole problem of Ref.~\citenum{moroz_recursive_2005} can be reproduced with machine accuracy. For a more realistic description of the experimental setup, a three-dimensional electrodynamic problem was developed in the CST Studio Suite by means of the frequency domain solver~\cite{CST}. 
Those simulations can also be tuned to yield numerically accurate solutions. Though, incremental changes in some geometrical parameters, e.g. in a configuration of the feeding cables, were found to cause significant changes in the S-parameters in some spectral regions. Given that assessing these parameters experimentally is very difficult, there could be relatively large uncertainties related to their values even for relatively simple models as considered here~\cite{Ant_meas}.
Experimental data, including finite calibration tolerances ($<1\%$), inaccuracies in the spatial orientation and position of the transmitting and receiving antennas ($<1\%$), polarization loss ($<0.2\%$), noise ($<0.1\%$), parasitic reflections and radiation from equipment, not ideal parameters of the cables, dipole, and dielectric resonator are the largest source of potential uncertainties and errors. Also, the directivity was evaluated with the S-parameters that were measured in only two planes ($\phi=0^{\circ}$ and $\phi=90^{\circ}$), which also bring  an error related to the realistic pattern deviations when rotating around the z-axis.

Surface irregularities and/or inhomogeneity of the sphere material lead to frequency splitting of excited modes with different magnetic number $m$, which leads to violation of the superdirectivity.
Superdirective antennas have been in theory  often considered with an infinite ground plane (see examples in Ref.~\citenum{kildal_degrees_2017}), which is  not taken into account when calculating the practical antenna size. However, in experimental realizations, this ground plane is comparable to the size of the antenna and can have a significant impact on the distribution of near fields.

Regarding practical realization of the directivity limit a number of issues have to be taken into account.
It is important to understand that the frequency bandwidth significantly  decreases in the limit $\varepsilon_1\rightarrow\infty$, which, together with the requirements on manufacturing quality, makes it impossible to use theoretically obtained resonances in practice. On the other hand, when $\varepsilon_1\rightarrow 1$, it is as a rule impossible to excite well separated electric and magnetic modes of a spherical resonator at a given frequency.
We believe that the value of $\varepsilon'_1=12.2$ used in our experiment is a reasonable compromise for demonstrating the superdirectivity effect when operating at the required modes. 

To design a super-directional antenna similar to the described one with the internal dipole, the following steps can be proposed. First, one should choose a sphere with a refractive index and size allowing for its TE$_{401}$ eigenmode to be at a desired operating frequency \big(see Eqs.~(\ref{eq:char_eq}) and (\ref{eq:char_eq_short})\big). Next, for the required $kR$, an approximate optimal dipole position is selected on the basis of point dipole approximation (see Fig.~\ref{fig:dirmap}). Finally, a full-scale simulation and fine-tuning of the dipole parameters is carried out in order to minimize the return loss at the resonant frequency.
The shape of the feed element can be further optimized to improve matching and reduce dependence on the position of the dipole inside the resonator.

Finally, it is important to mention the issue of the antenna bandwidth. This is an ambiguous parameter~\cite{Balanis2016}, specifically for electrically small antennas. Typically, such antennas have low total radiation efficiency, whereas in our measurements, the radiation efficiency exceeded $80\%$ within the $12$ MHz frequency band, and the corresponding directivity reached the values of $8-10$ for the case with an external dipole, and $6-9$ for the case with an internal dipole.

 \section{Conclusion}
 \label{sec:summary}
{The novelty of our work lies in that a spherical dielectric resonator can facilitate an antenna of a simple design having superdirectional radiation in a given direction.}
The use of high-index, low-loss ceramics allows one to significantly reduce the physical size of antennas, while maintaining a high overall radiation efficiency and practical bandwidth. We have demonstrated both numerically and experimentally the possibility of superdirectional behavior for antennas smaller than the wavelength based on a spherical dielectric resonator with a dipole source. The antenna directivity is maintained at the frequencies of eigenmodes of the spherical resonator, provided that one appropriately adjusts the position of the dipole source.
Our results can be extrapolated to a wide range of radio~\cite{Luk2003, Narang2010, Parshin2018, dielectrics, Ohsato18}, and optical~\cite{Kuznetsov2016, Smirnova2016, Baranov2017, Tzarouchis2018, Odit2021} frequencies with a suitable choice of high-index materials, while taking into account the cost, availability, and safety of the materials, as well as their compatibility with the desired manufacturing method. It should also be noted that reducing the size of the antenna leads to the main problem of all electrically small antennas, namely the problem of impedance matching. Unfortunately, this problem is fundamental and as the size decreases, the maximum achievable overall radiation efficiency of the antenna also decreases~\cite{harrington_effect_1960,Hansen2006,karlsson_efficiency_2013}. An ideal solution for this problem has not yet been obtained. Therefore, when the size of the antenna is greatly reduced relative to the wavelength, the maximum directivity can remain approximately the same, but the efficiency-dependent gain and the realized gain will decrease.

 \section*{SUPPLEMENTARY MATERIAL}
 \label{sec:suppl}
See the supplementary material for details related to the fabrication process of the antenna and more detailed data obtained during the measurements.

 \section*{ACKNOWLEDGMENT}
 \label{sec:Acknow}
Theoretical part of the work was supported by the Russian Science Foundation (Project No. 21-79-30038). Experimental part of the work was supported by the Priority 2030 Federal Academic Leadership Program.

\section*{AUTHOR DECLARATIONS}
\subsection*{Conflict of interest}
The authors declare that they have no competing interests.
\section*{DATA AVAILABILITY}
The data that support the findings of this study are available from the corresponding author upon reasonable request.

%

\end{document}



\title{Supplementary material for:\\
``Experimental demonstration of superdirective spherical dielectric antenna''}

\author{Roman Gaponenko,$^{1,*}$ Mikhail S. Sidorenko,$^{1}$ Dmitry Zhirihin,$^{1}$ Ilia L. Rasskazov,$^{2}$ Alexander Moroz,$^3$ Konstantin Ladutenko,$^1$ Pavel Belov,$^1$ and Alexey Shcherbakov$^{1}$}
\affiliation{%
 $^1$ School of Physics and Engineering, ITMO University, 197101, Saint-Petersburg, Russia \\ $^2$ KLA Corporation, 5 Technology Drive, Milpitas, California 95035, USA \\ $^3$ Wave-scattering.com \\
* Corresponding author: roman.gaponenko@metalab.ifmo.ru
}%

\maketitle

\clearpage
\setcounter{equation}{0}
\setcounter{figure}{0}
\setcounter{table}{0}
\renewcommand{\theequation}{S\arabic{equation}}
\renewcommand{\thefigure}{S\arabic{figure}}
\renewcommand{\thetable}{S\arabic{table}}
\onecolumngrid

\section{Preparing for measurements}
\label{sec:prep}

\begin{figure}[h!]
\begin{centering}
\includegraphics[width=0.95\textwidth]{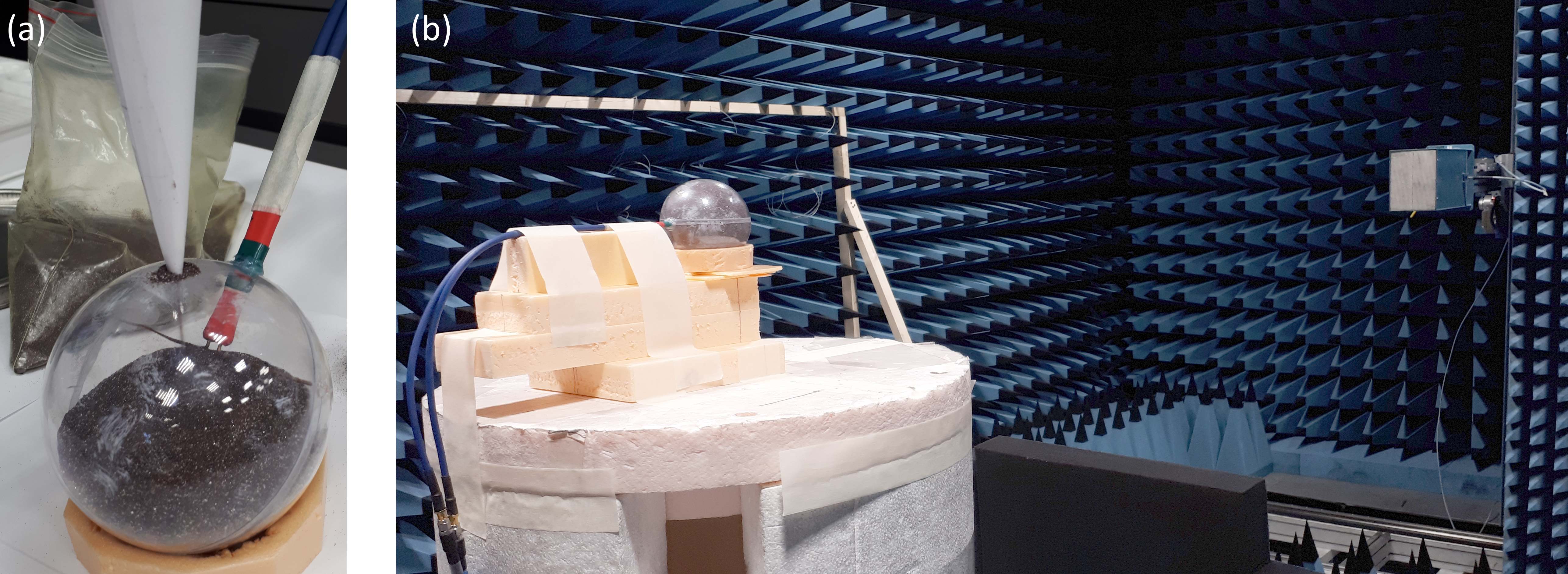}
\par\end{centering}
\caption{\label{fig:sph}Preparing for measurements. (a) Antenna fabrication process and (b) a full-view of the measurement setup.}
\end{figure}

Figure~\ref{fig:sph}(a) is a photograph of the fabrication process of a spherical dielectric antenna using Free Flowing Dielectric Powder ECCOSTOCK HiK. It has has the real part of permittivity $\varepsilon'_1\simeq 12.2$ and the loss tangent $\tan\delta_1\simeq 0.0007$. The powder was poured into a plastic shell made of acrylic with a thickness of $\sim1.1$ mm, the outer radius of which is $R_2=40$ mm, the real part of permittivity $\varepsilon'_2\simeq1.7$ and the loss tangent $\tan \delta_2\simeq0.02$. To achieve a homogeneous structure of the powder, it was compacted using vibrations. After the installation and antenna alignment process, the foam pedestal (on a rotary positioner) was rotated in increments of $1^\circ$ along the $\theta$ angle, with the measurement of S-parameters at each step. Figure~\ref{fig:sph}(b) is a photograph of an anechoic chamber, where the antenna under test is installed on the left on the special foam pedestal, and a horn antenna is on the right on the triaxial positioner. The secant plane $\phi=0^\circ$ includes the arms of the dipole and the center of the sphere. To take measurements in the $\phi=90^\circ$ plane (which is perpendicular to the dipole arms), the receiving and transmitting antennas were preliminarily rotated by $90^\circ$ around the axis on which both of these antennas are located. After that, a similar measurement of S-parameters was performed along the $\theta$ angle for the plane $\phi=90^\circ$.

\clearpage
\section{Experimental data}
\label{ap:sec2}

\subsection{Dielectric sphere excited by an external dipole}

\begin{figure}[h!]
\begin{centering}
\includegraphics[width=\textwidth]{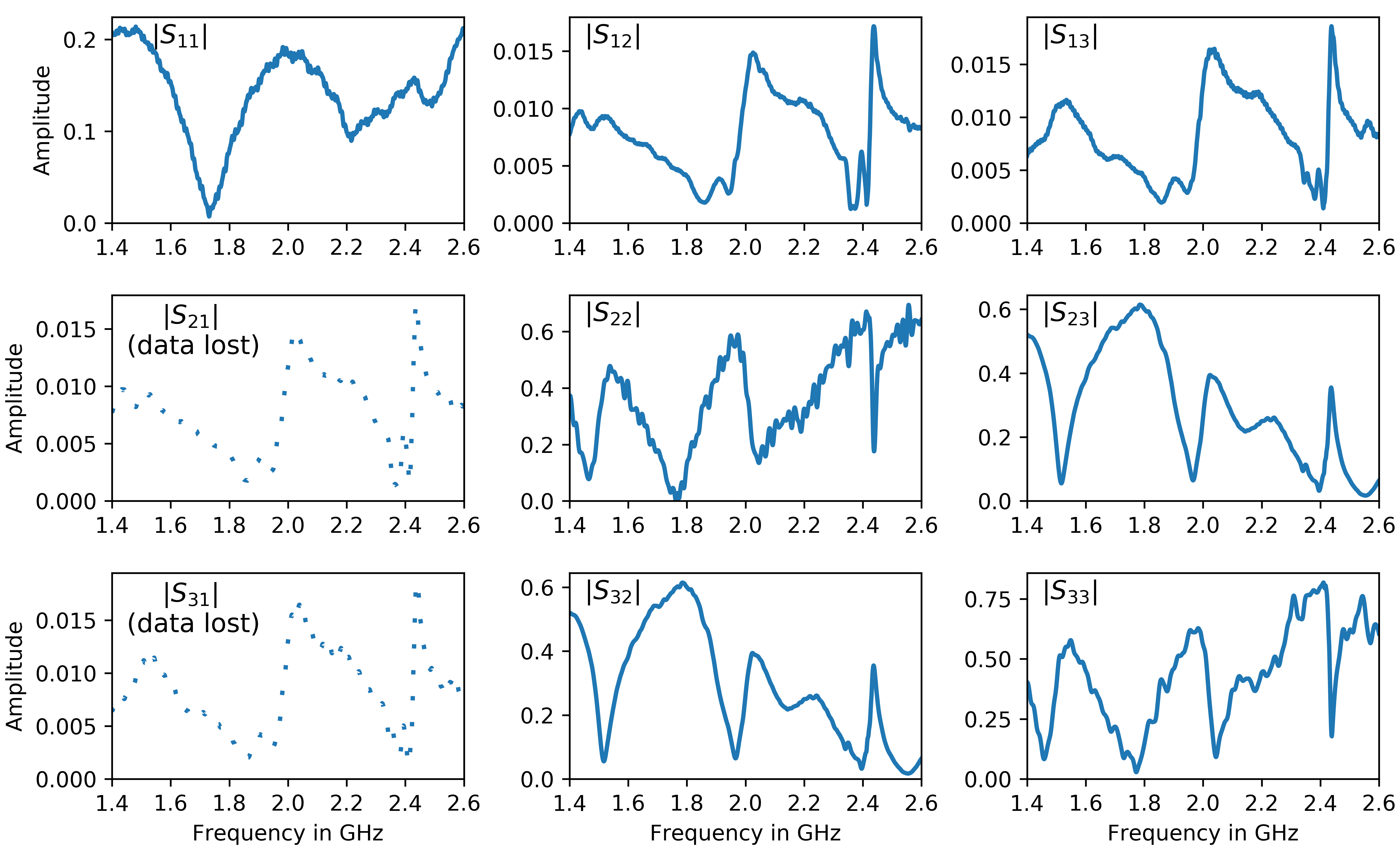}
\par\end{centering}
\caption{\label{fig:S_param_out1}Measured S-parameters for the dielectric spherical antenna excited by an external dipole. Antenna construction is described in Section~III~A of the main text of the article.}
\end{figure}

\begin{figure}[h!]
\begin{centering}
\includegraphics[width=\textwidth]{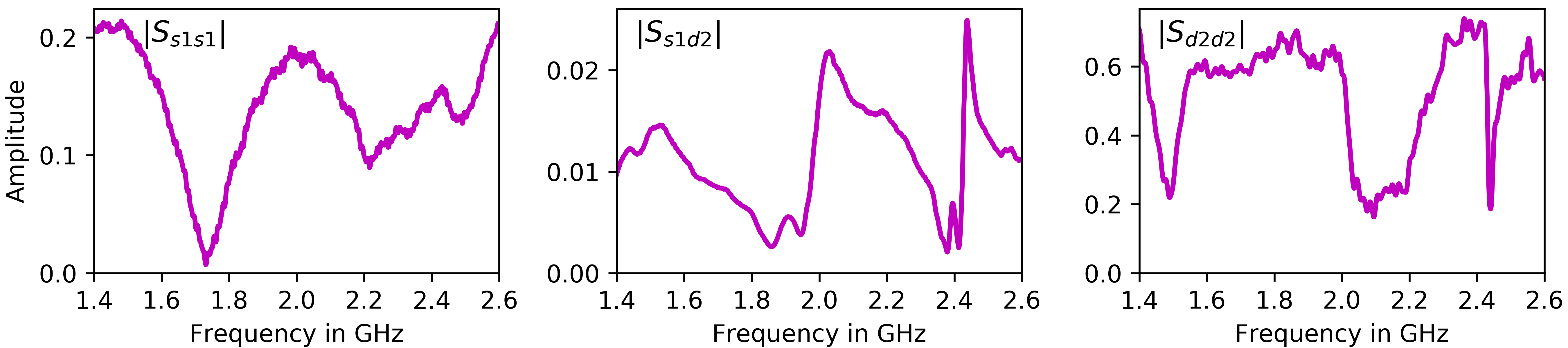}
\par\end{centering}
\caption{\label{fig:S_param_out2}Calculated S-parameters for the dielectric spherical antenna excited by an external dipole. Antenna construction is described in Section~III~A of the main text of the article.}
\end{figure}

\clearpage
\subsection{Dielectric sphere excited by an internal dipole}

\begin{figure}[h!]
\begin{centering}
\includegraphics[width=\textwidth]{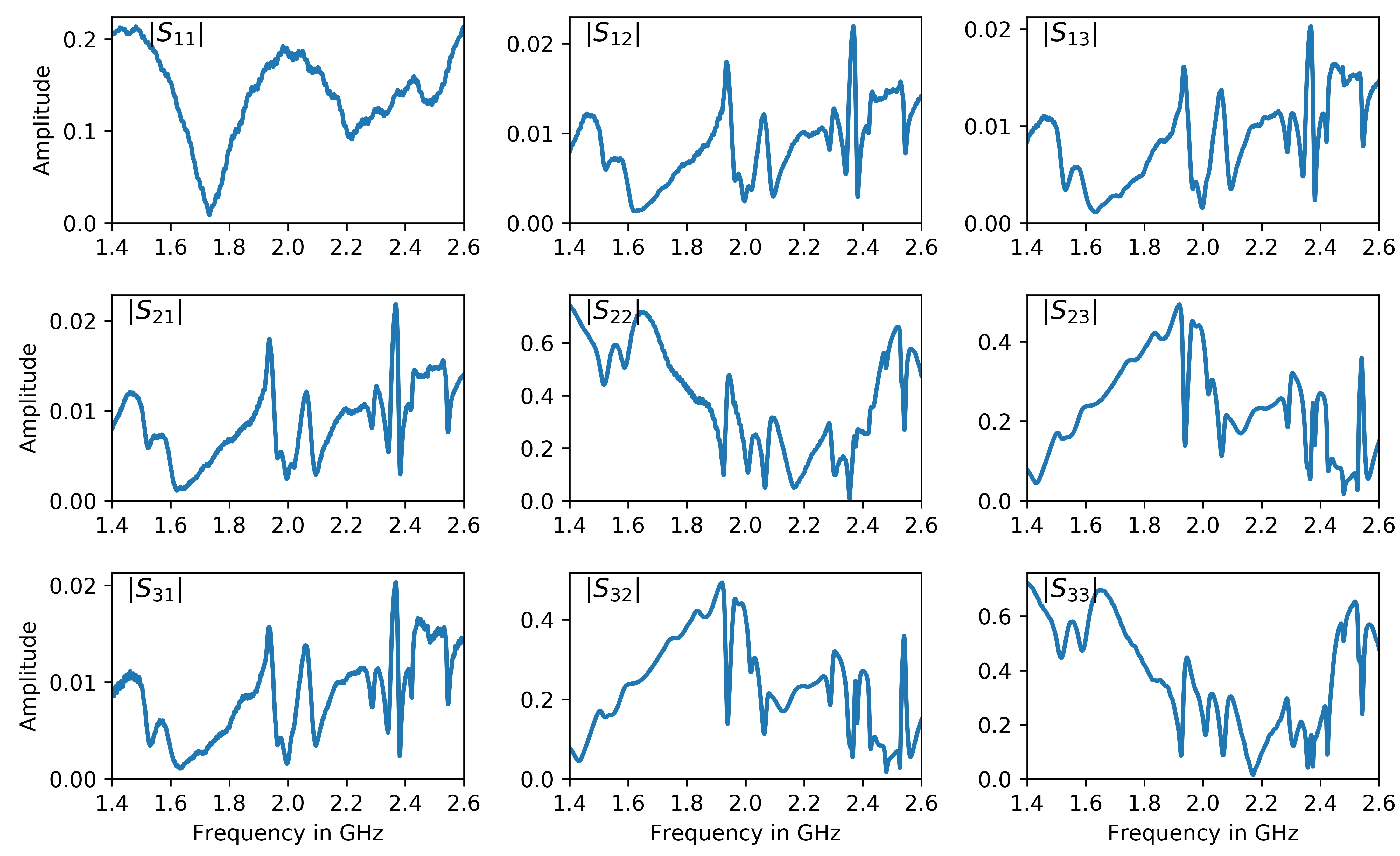}
\par\end{centering}
\caption{\label{fig:S_param_in1}Measured S-parameters for the dielectric spherical antenna excited by an internal dipole. Antenna construction is described in Section~III~B of the main text of the article.}
\end{figure}

\begin{figure}[h!]
\begin{centering}
\includegraphics[width=0.66\textwidth]{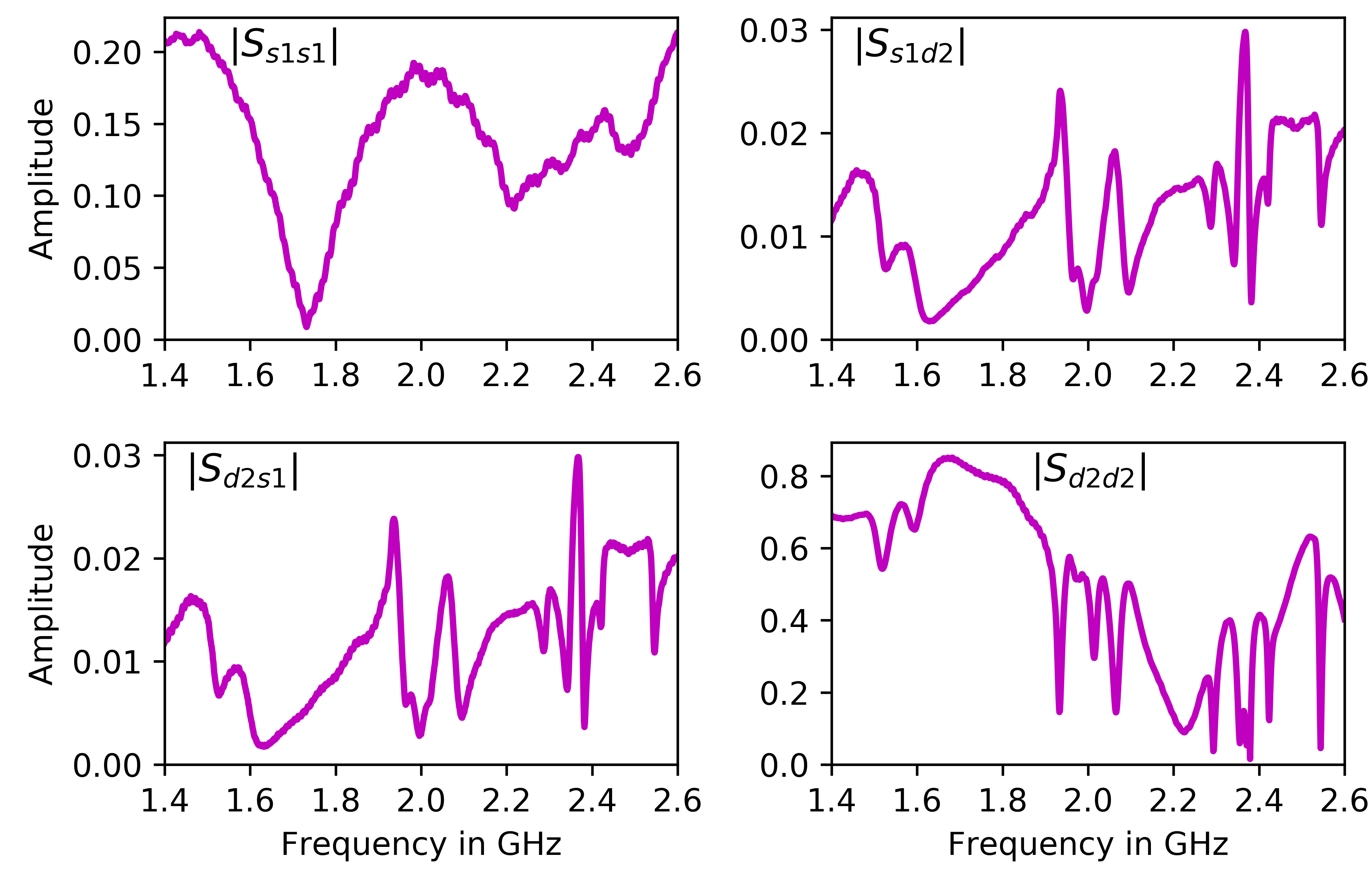}
\par\end{centering}
\caption{\label{fig:S_param_in2}Calculated S-parameters for the dielectric spherical antenna excited by an internal dipole. Antenna construction is described in Section~III~B of the main text of the article.}
\end{figure}


\section{Antenna efficiency}
\subsection{Dielectric sphere excited by an external dipole}
\begin{figure*}[h!]
\includegraphics[width=\textwidth]{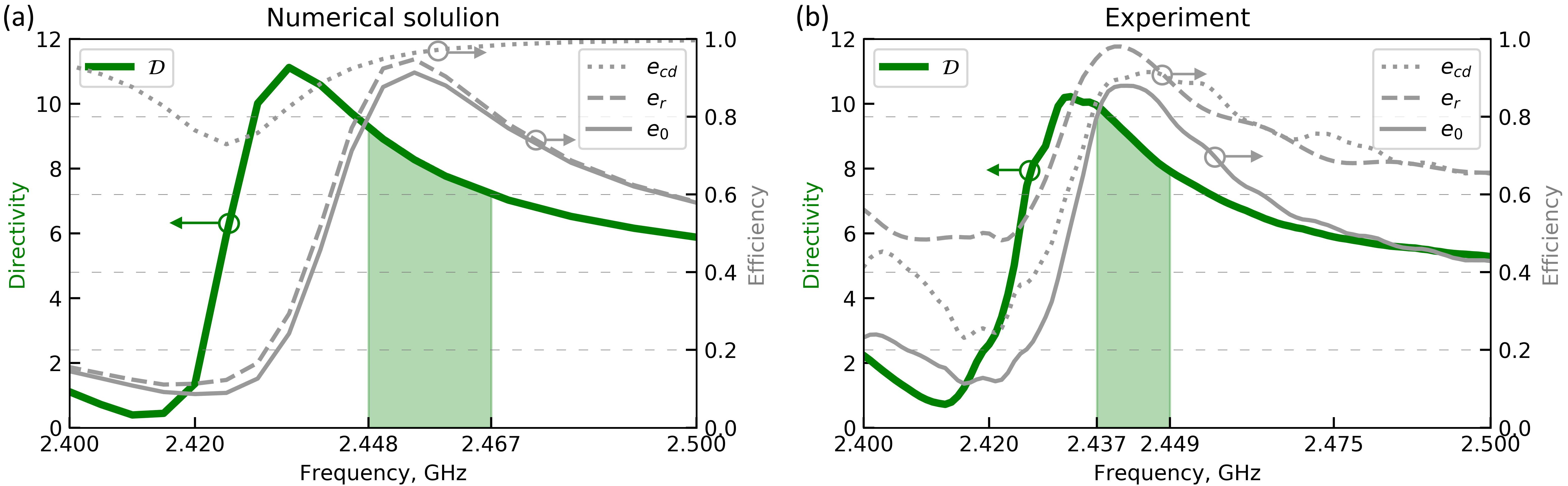}%
\caption{\label{fig:band_ext}Spectra of the directivity ($\mathcal{D}$) in the forward direction (along the $z$-axis) and efficiency in the range $2.4-2.5$ GHz obtained (a) numerically and (b) experimentally. Antenna construction is described in Section~III~A of the main text. The left panel shows the results of the numerical calculation performed in the CST Studio Suite for the case of a full-size dipole oriented tangentially and located at a distance of $r_d=40.5$ from the center of the spherical resonator with the radius of $40$ mm. The right panel shows the experimentally measured results.}
\end{figure*}

$e_{cd}=\mathcal{G}/\mathcal{D}$ is the antenna radiation efficiency (dimensionless);
\hspace{1.4cm}$\mathcal{D}$ is the antenna directivity;

$e_r=\mathcal{G}_{\cal R}/\mathcal{G}$ is the reflection (mismatch) efficiency (dimensionless);
\hspace{0.84cm}$\mathcal{G}$ is the antenna gain;

$e_0=\mathcal{G}_{\cal R}/\mathcal{D}=e_r e_{cd}$, is the total efficiency (dimensionless);
\hspace{2.0cm}$\mathcal{G}_{R}$ is the realized gain.

\subsection{Dielectric sphere excited by an internal dipole}
\begin{figure*}[h!]
\includegraphics[width=\textwidth]{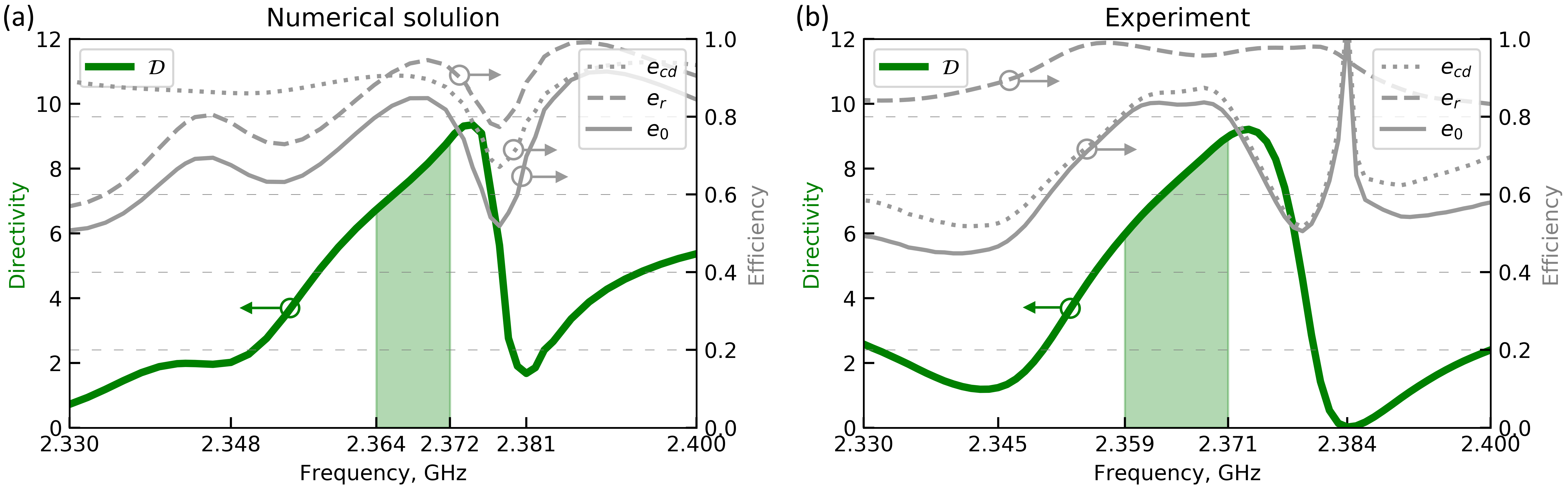}%
\caption{\label{fig:band_in}Spectra of the directivity ($\mathcal{D}$) in the forward direction (along the $z$-axis) and efficiency in the range $2.33-2.4$ GHz obtained (a) numerically and (b) experimentally. Antenna construction is described in Section~III~B of the main text. The left panel shows the results of the numerical calculation performed in the CST Studio Suite for the case of a full-size dipole oriented tangentially and located at a distance of $r_d=14.5$ from the center of the spherical resonator with the radius of $40$ mm. The right panel shows the experimentally measured results.}
\end{figure*}

Frequencies where the overall efficiency is $>80\%$ near the main resonance are marked with green shading. It should be noted that the accuracy of the experimental results could be affected by numerous reasons indicated in Section IV of the main text of the article. If the realized gain measurements are obtained by directly measuring the S-parameters, then the directivity measurements were calculated from the shape of the pattern in only two planes ($\phi=0^\circ$ and $\phi=90^\circ$). This could lead to some inaccuracies at frequencies outside the main resonances, including the zone near $\mathcal{D} \simeq 0$ and in the case of the appearance of side lobes outside the mentioned planes. The presented results of numerical optimization were obtained in the process of maximizing the realized gain, and if the goal was to maximize directivity, then for the same sphere and at the same resonances it is possible to obtain $\mathcal{D} \simeq 11$, but with a low total radiation efficiency.

\section{Influence of the plastic shell on the results of the experiment}

\begin{figure*}[h!]
\includegraphics[width=0.9\textwidth]{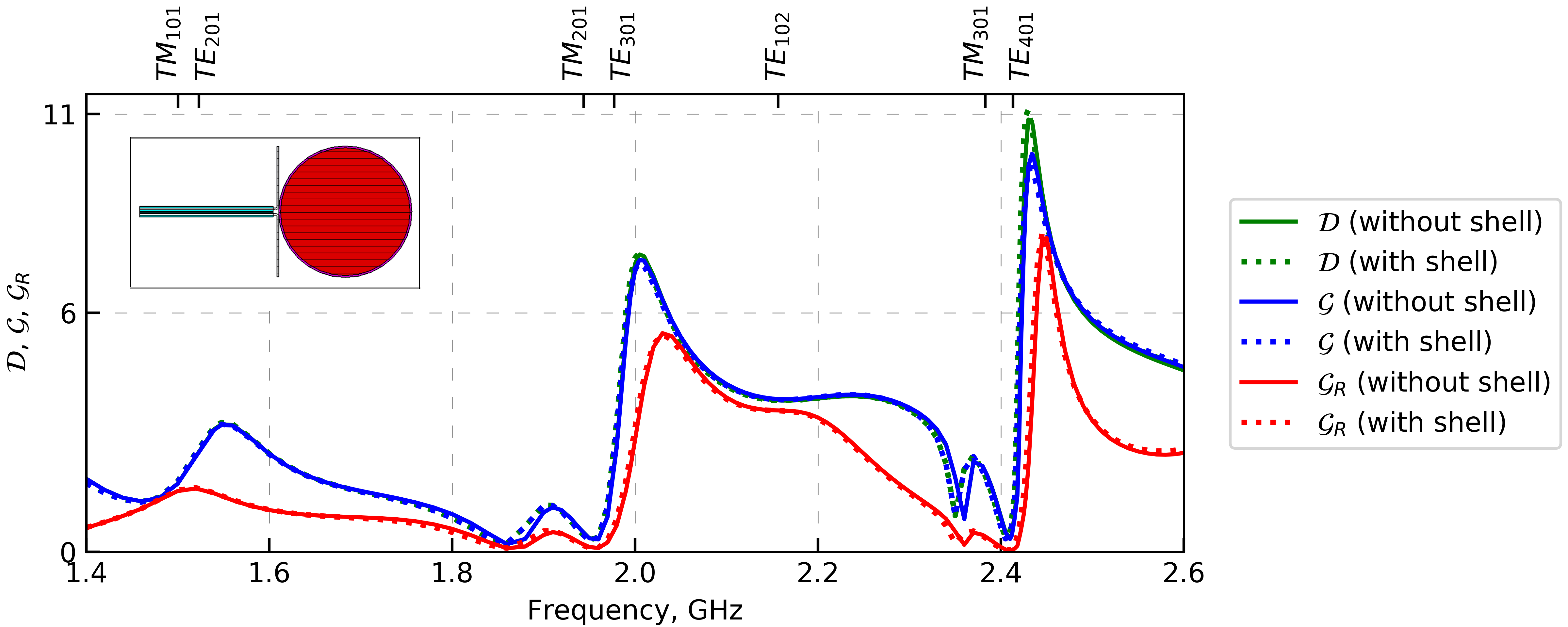}%
\caption{\label{fig:shell}Antenna characteristics ($\mathcal{D}$, $\mathcal{G}$, $\mathcal{G}_{R}$) in the forward direction (along the $z$ axis) in the range of $1.4-2.6$ GHz for the case a full size dipole oriented tangentially and spaced $r_d=40.5$ from the center of the spherical resonator as shown in the upper left corner. The outer shell of the resonator is made of plastic with an outer radius of $40$ mm, a thickness of $1.1$ mm, having $\varepsilon'_2=1.7$ and $\tan\delta_2=0.02$. The main body of the spherical resonator with a radius of $38.9$ mm is made of a dielectric material with $\varepsilon'_1=12.2$ and $\tan\delta_1=0.0007$. Other parameters of the considered antenna are described in Section~III~A. The results were obtained by numerical simulation in the CST Studio Suite for the case with and without a plastic shell.}
\end{figure*}

\footnotesize

%